\documentclass[intlimits,twoside,a4paper]{article}

\usepackage{amsmath,amssymb}
\usepackage{graphicx,color}

\usepackage[T2A]{fontenc}
\usepackage[cp1251]{inputenc}

\usepackage{cmpj2}

\issue{2012}{15}{4}{43101}

\doinumber{10.5488/CMP.15.43101}

\title[Some statistical aspects of the spinor field Fermi-Bose duality]%
{Some statistical aspects of the spinor field Fermi-Bose
duality}

\author[V.M. Simulik, I.Yu. Krivsky, I.L. Lamer]{V.M. Simulik\thanks{E-mail: vsimulik@gmail.com}\,, \ I.Yu. Krivsky, \ I.L. Lamer}
\address{Institute of Electron Physics of the National Academy of
Sciences of Ukraine, \\ 21 Universitetska St., 88000 Uzhgorod,
Ukraine}

\authorcopyright{V.M. Simulik, I.Yu. Krivsky, I.L. Lamer, 2012}
\date{Received June 25, 2012, in final form September 14, 2012}

\begin{document}

\maketitle

\begin{abstract}
The structure of 29-dimensional extended real Clifford-Dirac
algebra, which has been introduced in our paper Phys. Lett. A, 2011,
\textbf{375}, 2479, is considered in brief. Using this
algebra, the property of Fermi-Bose duality of the Dirac
equation with nonzero mass is proved. It means that Dirac
equation can describe not only the fermionic but also the bosonic
states. The proof of our assertion based on the examples of bosonic
symmetries, solutions and conservation laws is given. Some
statistical aspects of the spinor field Fermi-Bose duality are
discussed.


\pacs 11.30-z., 11.30.Cp., 11.30.j
\end{abstract}

\section{Introduction}

Some new statistical aspects of the Dirac equation are considered.
The Fermi-Bose (FB) duality of the spinor field has been originally
mentioned by L.~Foldy~\cite{1}. An extended consideration has
been given in~\cite{2}. P.~Garbaczewski proved~\cite{2} that the Fock space
$\mathcal{H}^{\mathrm{F}}(\mathrm{H}^{3,\mathrm{M}})$ over the
quantum mechanical space
$\mathrm{L}_{2}(\mathrm{R}^3)\otimes\mathrm{C}^{\otimes\mathrm{M}}$
of a particle, which is described by the field
$\phi:\mathrm{M}(1,\mathrm{N})\rightarrow\mathrm{C}^{\otimes\mathrm{N}}$,
allows one to fulfill the dual FB quantization of the field $\phi$ in
$\mathcal{H}^{\mathrm{F}}$. Both the canonical commutation
relations (CCR) and the anticommutation relations (CAR) were used to
realize the above mentioned quantization. Moreover, for both
types of quantization, the uniqueness of the vacuum in
$\mathcal{H}^{\mathrm{F}}$ was proved. The dual FB quantization
was illustrated by various examples and in the spaces
$\mathrm{M}(1,\mathrm{N})$ of arbitrary dimensions. The massless
spinor field was considered in detail~\cite{2}.

In our publications, the consideration of the FB duality
conception of the field was extended by applying the
group-theoretical approach to the problem (we often referred to the FB duality as the relationship between the fields of integer
and half-integer spins, see e.g.~\cite{3,4,5,6,7}). As the first step, we
have considered in detail the case of massless Dirac equation.
Both Fermi and Bose local representations of the universal
covering $\mathcal{P}\supset\mathcal{L}$ = SL(2,C) of proper
ortochronous Poincar\'{e} group $\mbox{P}_ + ^
\uparrow = \mbox{T(4)}\times )\mbox{L}_ + ^ \uparrow
\supset\mbox{L}_ + ^ \uparrow $ = SO(1,3), with respect to which
the Dirac equation is invariant, were found. The same was
realized~\cite{6} for the slightly generalized original Maxwell
equations, in which the complex valued 4-object
$\mathcal{E}(x)=E(x)-\ri H(x)$ of field strengths is the
tensor-scalar (s=1,0)  $\mathcal{P}$-covariant. In general, we
have proved the existence of bosonic symmetries, solutions and
conservation laws for a massless Dirac equation~\cite{3,4,5,6,7}. Thus,
a systematic investigation of the bosonic properties of a
massless Dirac equation was carried out.

In our investigations, we followed the authors of a number of
papers, in which they considered the problem of the relationship between
the Dirac and Maxwell equations starting from the very origin of quantum mechanics~\cite{8,9,10,11,12,13,14,15,16,17,18}. However, the authors
of these papers considered the simplest example of a free,
massless Dirac equation and its relation to the Maxwell
equations. Interest to such problems has grown after the
investigations~\cite{19,20} of a physically meaningful case (i.e., mass
is nonzero and the interaction potential is nonzero too) and following our
own research steps~\cite{21,22,23} in the same direction.
Unfortunately, only stationary Dirac and Maxwell equations
were considered.

In another approach~\cite{24,25,26,27,28,29,30}, the quadratic relations between
the fermionic and bosonic amplitudes were found and applied. In our
papers~\cite{3,4,5,6,7,21,22,23} and herein we discuss the linear relations
between fermionic and bosonic amplitudes.

Our results were further developed and employed by the authors of~\cite{31,32,33,34,35,36,37,38},
where the references to the above mentioned papers of ours were made.
Nevertheless, the general case, where the mass in the Dirac
equation is not equal to zero, is still open for investigations
and considerations.

Only recently~\cite{39,40,41,42} we were able to extend our consideration
to a Dirac equation with a nonzero mass. The important step was
as follows. We started with the Foldy-Wouthuysen (FW)~\cite{43}
representation of the Dirac equation, and the results for a
standard Dirac equation were found as a consequence of the FW
transformation. In our papers~\cite{39,40,41,42}, bosonic representations
of universal covering $\mathcal{L}$ = SL(2,C) of proper
ortochronous Lorentz group $\mbox{L}_ + ^ \uparrow $ = SO(1,3)
were found, with respect to which the Dirac and FW equations with
nonzero mass are invariant. The main results herein are the bosonic
spin (1,0) representations of Poincar\'{e} group
$\mathcal{P}$, with respect to which these equations are
invariant. These results were proved based on the  64
dimensional extended real Clifford-Dirac (ERCD) algebra and 29
dimensional proper ERCD algebra, which were taken into
consideration in~\cite{39,40,41,42} and essentially generalized the
standard 16 dimensional Clifford-Dirac (CD) algebra.

Here we consider (i) the dual (fermionic and bosonic) symmetries~\cite{39,40,41,42}
of the FW~\cite{43} and Dirac equations with nonzero mass,
(ii) continue the construction of bosonic solutions~\cite{44} of these
equations and (iii) demonstrate the existence of both Fermi and
Bose conservation laws for a spinor field. Thus, we present the
third level proof of the FB duality of the Dirac equation.
Moreover, the statistical analysis of the Dirac model of the spin
1/2 particle doublet description is presented. We prove that  such an
analysis can be successful only if one uses the standard
quantum-mechanical probability amplitudes distribution with
respect to the eigenvalues of complete sets of experimental
observable quantum-mechanical physical values. We presented in
detail the corresponding quantum-mechanical stationary complete
sets of operators of FB physical quantities. This allows us to
demonstrate the statistical aspect of the spinor field FB duality.

For our purposes we use the mathematical formalism of the ERCD
algebra and proper ERCD algebra~\cite{39,40,41,42}.

In section~2, the necessary notations and definitions are presented.

In section~3, the 29-dimensional proper extended real Clifford-Dirac algebra~\cite{40,41,42}, which is the mathematical basis of our
consideration, is presented in brief.

In section~4, the bosonic spin $s=(1,0)$ symmetry~\cite{40,41,42} of the
FW and Dirac equations with nonzero mass is briefly considered as
the first step in our proof of the Dirac equation FB duality.

In section~5, we continue to construct the bosonic~\cite{44} spin
$s=(1,0)$ multiplet solutions of the FW and Dirac equations with
nonzero mass. This is the second step in our proof of the Dirac
equation FB duality.

In section~6, the FB duality of the spinor field is demonstrated
based on the example of the existence of both Fermi and Bose series of
conservation laws for this field (i.e., the third step of our proof).

In section~7, the statistical aspects of the spinor field FB
duality are considered.

In section~8, brief general conclusions are formulated.

\section{Notations and definitions}

The system of units $\hbar=c=1$ and metric
$g=(g^{\mu\nu})=(+---)$, $a^{\mu}=g^{\mu\nu}a_{\nu}$, are taken.
The Greek indices are changed in the region
$0,1,2,3\equiv\overline{0,3}$, Latin~--- $\overline{1,3}$, the
summation over the twice repeated index is implied. The Dirac
$\gamma^{\mu}$ matrices in the standard Pauli-Dirac (PD)
representation are used. Our consideration is fulfilled in the
rigged Hilbert space
$\mathrm{S}^{3,4}\subset\mathrm{H}^{3,4}\subset\mathrm{S}^{3,4*}$,
where $\mathrm{H}^{3,4}$ is given by
\begin{equation}
\label{eq1}
\mathrm{H}^{3,4}=\mathrm{L}_{2}(\mathrm{R}^3)\otimes\mathrm{C}^{\otimes4}=\left\{\phi=(\phi^{\mu}):\mathrm{R}^{3}\rightarrow\mathrm{C}^{\otimes4};
 \, \int
\rd^{3}x|f(t,\vec{x})|^{2} <\infty\right\}
\end{equation}
and symbol `$\ast$' in $\mathrm{S}^{3,4*}$ means, that the
space of Schwartz generalized functions $\mathrm{S}^{3,4*}$ is
conjugated to the Schwartz test function space $\mathrm{S}^{3,4}$
by the corresponding topology. For more details see~\cite{41}.

We consider the ordinary CD algebra to be the algebra of $4\times4$
Dirac matrices in the standard PD representation in terms of the
standard $2\times2$ Pauli matrices.

For the purposes dealing with physics it is useful to consider
the corresponding groups an algebras with real parameters (e.g.,
the parameters $a=(a^{\mu}),\, \omega=(\omega^{\mu\nu})$ of the
translations and rotations for the group $\mbox{P}_ + ^
\uparrow$). Therefore, corresponding generators are
anti-Hermitian. The mathematical correctness of such a choice of
generators is verified in~\cite{45,46}.

\section{Proper extended real Clifford-Dirac algebra}

We consider the standard 16-dimensional CD algebra of the
matrices to be a real one and add the imaginary unit $\ri=\sqrt{-1}$
together with the operator $\hat{C}$ of complex conjugation (the
involution operator in the space $\mathrm{H}^{3,4}$) into the set
of the CD algebra possible generators. This enabled us to extend
the standard CD algebra up to the 64-dimensional extended real CD
algebra (i.e., ERCD algebra of~\cite{40,41,42}). Here, the subalgebras of the
ERCD algebra are considered in brief. The most important are the
representations in $\mathrm{C}^{\otimes4}\subset\mathrm{H}^{3,4}$
of the 29-dimensional proper ERCD algebra SO(8) spanned on the
orts
\begin{equation}
\label{eq2}
\gamma^{1},\qquad\gamma^{2},\qquad\gamma^{3},\qquad\gamma^{4}=\gamma^{0}\gamma^{1}\gamma^{2}\gamma^{3},
\qquad\gamma^{5}=\gamma^{1}\gamma^{3}\hat{C},
\qquad\gamma^{6}=\ri\gamma^{1}\gamma^{3}\hat{C},\qquad\gamma^{7}=\ri\gamma^{0},
\end{equation}
where $ \gamma ^0 = \left| {{\begin{array}{*{20}c}
 1 \hfill & 0 \hfill \\
 0 \hfill & { - 1} \hfill \\
\end{array} }} \right|$,  $\gamma ^k = \left| {{\begin{array}{*{20}c}
 0 \hfill & {\sigma ^k} \hfill \\
 { - \sigma ^k} \hfill & 0 \hfill \\
\end{array} }}\right|$ and $\sigma ^k$ are the standard Pauli
matrices. The generators (\ref{eq2}) satisfy the anticommutation
relations~\cite{1}
\begin{equation}
\label{eq3} \gamma ^\mathrm{A} \gamma ^\mathrm{B} + \gamma
^\mathrm{B}\gamma ^\mathrm{A} = -
2\delta^{\mathrm{A}\mathrm{B}}, \qquad \mathrm{A},\mathrm{B}=\overline{1,7},
\end{equation}
and the generators of the proper ERCD algebra
$\alpha^{\widetilde{\mathrm{A}}\widetilde{\mathrm{B}}}=2s^{\widetilde{\mathrm{A}}\widetilde{\mathrm{B}}}$
(together with the unit ort, $4\times4$ matrix $\mathrm{I}_{4}$, we
have 29 independent orts
$\mathrm{I}_{4},\,\alpha^{\widetilde{\mathrm{A}}\widetilde{\mathrm{B}}}
=2s^{\widetilde{\mathrm{A}}\widetilde{\mathrm{B}}}$)
\begin{equation}
\label{eq4}
s^{\widetilde{\mathrm{A}}\widetilde{\mathrm{B}}}=\left\{s^{\mathrm{A}\mathrm{B}}=\frac{1}{4}
\left[\gamma
^\mathrm{A},\gamma
^\mathrm{B}\right], \ s^{\mathrm{A}8}=-s^{8\mathrm{A}}=\frac{1}{2}\gamma
^\mathrm{A}\right\},\qquad \widetilde{\mathrm{A}},\widetilde{\mathrm{B}}=\overline{1,8}
\end{equation}
satisfy the commutation relations of SO(8) algebra
\begin{equation}
\label{eq5}
\left[s^{\widetilde{\mathrm{A}}\widetilde{\mathrm{B}}},s^{\widetilde{\mathrm{C}}\widetilde{\mathrm{D}}}\right]=
\delta^{\widetilde{\mathrm{A}}\widetilde{\mathrm{C}}}s^{\widetilde{\mathrm{B}}\widetilde{\mathrm{D}}}
+\delta^{\widetilde{\mathrm{C}}\widetilde{\mathrm{B}}}s^{\widetilde{\mathrm{D}}\widetilde{\mathrm{A}}}
+\delta^{\widetilde{\mathrm{B}}\widetilde{\mathrm{D}}}s^{\widetilde{\mathrm{A}}\widetilde{\mathrm{C}}}
+\delta^{\widetilde{\mathrm{D}}\widetilde{\mathrm{A}}}s^{\widetilde{\mathrm{C}}\widetilde{\mathrm{B}}}.
\end{equation}

In particular, the proper ERCD algebra SO(8), given by the 29 orts~(\ref{eq4}),
is our~\cite{40,41,42} direct generalization of the standard
16-dimensional CD algebra. It is also the basis for our dual FB
consideration of a spinor field, which enabled us to prove the
additional bosonic properties of this field. For physical
applications, we consider the realizations of the proper ERCD
algebra in the field space
$\mathrm{S}^{*}(\mathrm{M}(1,3))\otimes\mathrm{C}^{\otimes4}\equiv\mathrm{S}^{4,4*}$
of the Schwartz generalized functions and in the quantum
mechanical Hilbert space $\mathrm{H}^{3,4}$~(\ref{eq1}). These
realizations are found with the help of transformations
$V^{+}\mathrm{S}\mathrm{O}(8)V^{-},\,v\mathrm{S}\mathrm{O}(8)v$,
where the operators of transformations have the following form
\begin{equation}
\label{eq6}
V^{\pm}\equiv\frac{\pm \ri\vec{\gamma }\cdot \vec{\nabla}+\widehat{\omega }+m
}{\sqrt{2\widehat{\omega }(\widehat{\omega }+m)}}\, ,
\quad %
v=\left|
{{\begin{array}{*{20}c}
 \mathrm{I}_{2} \hfill & 0 \hfill \\
 0 \hfill & { \hat{C}\mathrm{I}_{2}} \hfill \\
\end{array} }} \right|,
\quad %
\widehat{\omega} \equiv \sqrt { - \Delta + m^2},
\quad %
\vec\nabla\equiv(\partial_{\ell}),
\quad %
\mathrm{I}_{2}=\left|
{{\begin{array}{*{20}c}
 1 \hfill & 0 \hfill \\
 0 \hfill & { 1} \hfill \\
\end{array} }} \right|. \\
\end{equation}
Furthermore, the realizations of the proper ERCD algebra
for bosonic fields are presented.

We take into consideration the ERCD algebra (64 orts) and the proper ERCD
algebra (29 orts) into the FW representation of the spinor field~\cite{43} (the advantages in comparison with the standard Dirac equation in
definitions of coordinate, velocity and spin operators are well
known from~\cite{43}). In this representation, the equation for the
spinor field (the FW equation) has the following form
\begin{equation}
\label{eq7} \left(\partial _0 + \ri\gamma^{0}\widehat{\omega} \right)\phi (x)
= 0, \qquad x\in \mathrm{M}(1,3), \qquad \phi\in \mathrm{H}^{3,4};
\end{equation}
and is linked with the Dirac equation
 \begin{equation}
\label{eq8}\left(\partial_{0}+\ri H\right)\psi(x)=0, %
\qquad%
H\equiv\vec{\alpha}\cdot\vec{p}+\beta
m; \qquad \vec{\alpha}\equiv\gamma^0\vec\gamma, \qquad  \beta\equiv\gamma^0
\end{equation}
by the FW transformation $V^{\pm}$
\begin{equation}
\label{eq9} \phi(x)=V^{-}\psi(x), \qquad \psi(x)=V^{+}\phi(x), \qquad
V^{+}\gamma^{0}\widehat{\omega
}V^{-}=\vec{\alpha}\cdot\vec{p}+\beta m.
\end{equation}
Herein below, the ERCD algebra and the proper ERCD algebra~(\ref{eq4}) are
essentially used in our proofs of bosonic properties of the Dirac
and FW equations. The proper ERCD algebra has 29 independent orts
presented in~(\ref{eq4}). In comparison with 16 independent orts of standard
CD algebra, we can operate now with additional elements. These
additional generators of SO(8) algebra enabled us to prove the
additional bosonic symmetries of the FW and Dirac equations~\cite{39,40,41,42} and to construct additional bosonic solutions of these
equations (\cite{44} and section~5 below). Moreover, the
anticommutation relations~(\ref{eq3}) were used in calculations.

\section{Bosonic spin $s=(1,0)$ symmetry of the Foldy-Wouthuysen and Dirac equations}

An example of the construction of an important bosonic symmetry
of the FW and Dirac equation is under consideration. A
fundamental assertion is that subalgebra SO(6) of the proper ERCD
algebra~(\ref{eq4}), which is determined by the operators
\begin{equation}
\label{eq10} \left\{\mathrm{I}, \,
\alpha^{\bar{\mathrm{A}}\bar{\mathrm{B}}}=2s^{\bar{\mathrm{A}}\bar{\mathrm{B}}}\right\},
\qquad \bar{\mathrm{A}},\bar{\mathrm{B}}=\overline{1,6},
\end{equation}
\begin{equation}
\label{eq11}\left\{s^{\bar{\mathrm{A}}\bar{\mathrm{B}}}\right\}
=\left\{s^{\bar{\mathrm{A}}\bar{\mathrm{B}}}\equiv\frac{1}{4}
\left[\gamma^{\bar{\mathrm{A}}},\gamma^{\bar{\mathrm{B}}}\right]\right\}
\end{equation}
is the algebra of invariance of the Dirac equation in
the FW representation~(\ref{eq7}) (in~(\ref{eq11}) the six matrices
$\{\gamma^{\bar{\mathrm{A}}}\}=\{\gamma^{1},\gamma^{2},\gamma^{3},\gamma^{4},\gamma^{5},\gamma^{6},\}$
are known from~(\ref{eq2})). Algebra SO(6) contains two different
realizations of SU(2) algebra for the spin s=1/2 doublet. By
taking the sum of the two independent sets of SU(2) generators
from~(\ref{eq11}), one can obtain the SU(2) generators of spin $s=(1,0)$
multiplet, which generate the transformation of the invariance of the
FW equation~(\ref{eq7}). These operators can be presented in the following form
\begin{equation}
\label{eq12}\vec{\breve{s}}\equiv\left(\breve{s}^{j}\right)=\left(\breve{s}_{mn}\right)
=\frac{1}{2}\left(\breve{\gamma}^{2}\breve{\gamma}^{3}-\breve{\gamma}^{0}\breve{\gamma}^{2}\breve{C},
\,\breve{\gamma}^{3}\breve{\gamma}^{1}+\breve{i}\breve{\gamma}^{0}\breve{\gamma}^{2}\breve{C},
\,\breve{\gamma}^{1}\breve{\gamma}^{2}-\breve{i}\right),
\end{equation}
where the corresponding orts of the ERCD algebra in bosonic
representation are given by
\begin{eqnarray}
\label{eq13}
&&\breve{\gamma}^{0} = \left| {{\begin{array}{*{20}c}
 \sigma ^3 \hfill & 0 \hfill \\
 0 \hfill & {\sigma ^1} \hfill \\
\end{array} }} \right|,
\qquad \breve{\gamma}^{1}=\frac{1}{\sqrt{2}}\left|
\begin{array}{cccc}
 0 & 0 & 1 & -1\\
 0 & 0 & \ri & \ri\\
-1 & \ri & 0 & 0\\
1 & \ri & 0 & 0\\
\end{array} \right|,
\qquad \breve{\gamma}^{2}=\frac{1}{\sqrt{2}}\left|
\begin{array}{cccc}
 0 & 0 & -\ri & \ri\\
 0 & 0 & -1 & -1\\
-\ri & 1 & 0 & 0\\
\ri & 1 & 0 & 0\\
\end{array} \right|,  \nonumber \\
%
%
&&\breve{\gamma}^{3} = - \left| {{\begin{array}{*{20}c}
 \sigma ^2 \hfill & 0 \hfill \\
 0 \hfill & {\ri\sigma ^2} \hfill \\
\end{array} }} \right|\hat{C},
\qquad
 \breve{\gamma}^{4} = \left| {{\begin{array}{*{20}c}
 \ri\sigma ^2 \hfill & 0 \hfill \\
 0 \hfill & {-\sigma ^2} \hfill \\
\end{array} }} \right|\hat{C},
\qquad \breve{\gamma}^{5}= \frac{1}{\sqrt{2}}\left|
\begin{array}{cccc}
 0 & 0 & -1 & -1\\
 0 & 0 & \ri & -\ri\\
1 & \ri & 0 & 0\\
1 & -\ri & 0 & 0\\
\end{array} \right|, \nonumber\\
%
&&\breve{\gamma}^{6}= \frac{1}{\sqrt{2}}\left|
\begin{array}{cccc}
 0 & 0 & -\ri & -\ri\\
 0 & 0 & 1 & -1\\
-\ri & -1 & 0 & 0\\
-\ri & 1 & 0 & 0\\
\end{array} \right|,
\quad
\breve{\gamma}^{7}=\gamma^{7}=\ri\gamma^{0},
\quad
\breve{i} = \left| {{\begin{array}{*{20}c}
 \ri\sigma ^3 \hfill & 0 \hfill \\
 0 \hfill & {-\ri\sigma ^1} \hfill \\
\end{array} }} \right|,
\quad \breve{C} = \left| {{\begin{array}{*{20}c}
 \sigma ^3 \hfill & 0 \hfill \\
 0 \hfill & {\mathrm{I}_{2}} \hfill \\
\end{array} }} \right|\hat{C}.
\end{eqnarray}
The spin operators~(\ref{eq12}) of SU(2) algebra, which commute
with the operator $\partial _0 + \ri\gamma^{0}\widehat{\omega }$ of
the FW equation~(\ref{eq7}), can also be presented in an explicit form
\begin{equation}
\label{eq14} \breve{s}^{1}= \frac{1}{\sqrt{2}}\left|
\begin{array}{cccc}
 0 & 0 & \ri\hat{C} & 0\\
 0 & 0 & -\hat{C} & 0\\
-\ri\hat{C} & \hat{C} & 0 & 0\\
0 & 0 & 0 & 0\\
\end{array} \right|, \qquad \breve{s}^{2}= \frac{1}{\sqrt{2}}\left|
\begin{array}{cccc}
 0 & 0 & \hat{C} & 0\\
 0 & 0 & -\ri\hat{C} & 0\\
-\hat{C} & \ri\hat{C} & 0 & 0\\
0 & 0 & 0 & 0\\
\end{array} \right|, \qquad \breve{s}^{3}= \left|
\begin{array}{cccc}
 -\ri & 0 & 0 & 0\\
 0 & \ri & 0 & 0\\
0 & 0 & 0 & 0\\
0 & 0 & 0 & 0\\
\end{array} \right|.
\end{equation}
The calculation of the Casimir operator for the SU(2)
generators~(\ref{eq14}) gives the following result
\begin{eqnarray*}
\vec{\breve{s}}^{2}
= -1(1+1)\times\left| \begin{array}{cc}
 \mathrm{I}_{3}  & 0  \\
 0  & {0}  \\
\end{array}  \right|.
\end{eqnarray*}

Transition from the fundamental representation A of the ERCD
algebra to the bosonic representation B is fulfilled
$\mathrm{B}=W\mathrm{A}W^{-1}$ using the operator $W$:
\begin{equation}
\label{eq15} W = \frac{1}{\sqrt{2}}\left|
\begin{array}{cccc}
 \sqrt{2} & 0 & 0 & 0\\
 0 & 0 & \ri\sqrt{2}\hat{C} & 0\\
0 & -\hat{C} & 0 & 1\\
0 & -\hat{C} & 0 & -1\\
\end{array} \right|, \quad W^{-1}=\left|
\begin{array}{cccc}
 \sqrt{2} & 0 & 0 & 0\\
 0 & 0 & -\hat{C} & -\hat{C}\\
0 & \ri\sqrt{2}\hat{C} & 0 & 0\\
0 & 0 & 1 & -1\\
\end{array} \right|, \quad WW^{-1}=W^{-1}W=\mathrm{I}_{4}\,.
\end{equation}

Based on the spin operators~(\ref{eq12}), (\ref{eq14}), the bosonic spin
(1,0) representation of the Poincar\'{e} group
$\mathcal{P}$ is constructed. It is easy to show (after our
consideration in~\cite{43} and above) that generators
\begin{equation}
\label{eq16}p_{0} = - \ri\gamma _{0}\widehat{\omega}, \quad
p_{n}=\partial _n\, , \quad
j_{ln}= x_l
\partial _n -
x_n \partial _l + \breve{s}_{ln}\,,\quad
j_{0k}=x_0\partial
_k+\ri\gamma_{0}\left\{x_k\widehat{\omega }+\frac{\partial
_k}{2\widehat{\omega }} +\frac{(\vec{\breve{s}}\times
\vec{\partial })_k}{\widehat{\omega }+m}\right\}
\end{equation}
of group $\mathcal{P}$ commute with the operator of the
FW equation~(\ref{eq7}) and satisfy the commutation relations of the Lie
algebra of the group $\mathcal{P}$ in a manifestly covariant form.
In the space $\mathrm{H}^{3,4}$, the operators~(\ref{eq16}) generate
a unitary $\mathcal{P}$ representation differing from the fermionic $\mathcal{P}^{\mathrm{F}}$-generators according to equations (D-64)--(D-67) in~\cite{1}, i.e.,
the bosonic $\mathcal{P}^{\mathrm{B}}$ representation of the
group $\mathcal{P}$, with respect to which the FW equation~(\ref{eq7}) is
invariant. For the generators~(\ref{eq16}), the Casimir operators have the
following form:
\begin{equation}
\label{eq17}p^{\mu}p_{\mu}=m^{2}, \qquad
W^{\mathrm{B}}=w^{\mu}w_{\mu}=m^{2}\vec{\breve{s}}^{2}=
-1(1+1)m^{2}\left| {{\begin{array}{*{20}c}
 \mathrm{I}_{3} \hfill & 0 \hfill \\
 0 \hfill & {0} \hfill \\
\end{array} }} \right|.
\end{equation}
Hence, according to the Bargman-Wigner classification, here
we consider  the spin $s=(1,0)$ representation of the group
$\mathcal{P}$.

The corresponding bosonic spin $s=(1,0)$ symmetries of the Dirac
equation~(\ref{eq8}) can be found from the generators~(\ref{eq16}) using the FW operator~(\ref{eq6}) in bosonic representation, i.e.,
$WV^{\pm}W^{-1}$.

More complete and detailed consideration of the bosonic
symmetries of the FW and Dirac equation was given in~\cite{40,41,42}.

\section{Bosonic spin $s=(1,0)$ multiplet solution of the Foldy-Wouthuysen and Dirac equations}

Here, as the next step in FB duality investigation, we consider the
bosonic solution of the Dirac (FW) equation. A bosonic solution
of the FW equation~(\ref{eq7}) is found completely similarly to the
procedure of construction of standard fermionic solution. Thus,
the bosonic solution is determined by some stationary diagonal
complete set of operators of bosonic physical quantities for the
spin $s=(1,0)$-multiplet in the FW representation, e.g., by the
set ``momentum-spin projection $\breve{s}^{3}$'':
\begin{equation}
\label{eq18} \left(\vec{p}=-\vec\nabla, \quad \breve{s}^{3}\right),
\end{equation}
where the spin operators $\vec{\breve{s}}$
and $\breve{s}^{3}$ for the spin $s=(1,0)$-multiplet are given in~(\ref{eq12}), (\ref{eq14}). The fundamental solutions of equation~(\ref{eq7}), which are
the common eigensolutions of the bosonic complete set~(\ref{eq18}), have
the following form
\begin{equation}
\label{eq19}
\varphi_{\vec{k}\mathrm{r}}^{-}(t,\vec{x})=\frac{1}{(2\pi)^{{3}/{2}}}\re^{-\ri kx}\mathrm{d}_{\mathrm{r}}\,,
\qquad \varphi_{\vec{k}\acute{\mathrm{r}}}^{+}(t,\vec{x})=\frac{1}{(2\pi)^{{3}/{2}}}\re^{\ri kx}\mathrm{d}_{\acute{\mathrm{r}}}\,,
\qquad kx=\omega t - \vec{k}\cdot\vec{x}\,,
\end{equation}
where $\mathrm{d}_{\alpha}=(\delta_{\alpha}^{\beta})$
are the Cartesian orts in the space
$\mathrm{C}^{\otimes4}\subset\mathrm{H}^{3,4}$, numbers
$\mathrm{r}=(1,2), \, \acute{\mathrm{r}}=(3,4)$ mark the eigenvalues $(+1,-1,0,\underline{0})$ of the operator $\breve{s}^{3}$
from~(\ref{eq12}), (\ref{eq14}).

The bosonic solutions of equation~(\ref{eq7}) are the generalized states
belonging to the space $\mathrm{S}^{3,4*}$; they form a complete
orthonormalized system of bosonic states. Therefore, any bosonic
physical state of the FW field $\phi$ from the dense in
$\mathrm{H}^{3,4}$ manifold $\mathrm{S}^{3,4}$ (the general
bosonic solution of the equation~(\ref{eq7})) is uniquely presented in
the following form
\begin{equation}
\label{eq20}\phi_{(1,0)}(x)=\frac{1}{(2\pi)^{{3}/{2}}}\int
\rd^{3}k\left[\xi^{\mathrm{r}}(\vec{k})\mathrm{d}_{\mathrm{r}}\re^{-\ri kx}
+
\xi^{*\acute{\mathrm{r}}}(\vec{k})\mathrm{d}_{\acute{\mathrm{r}}}\re^{\ri kx}\right],
\end{equation}
where $\xi (\vec{k})$ are the coefficients
of the expansion of bosonic solution of the FW equation~(\ref{eq7}) with
respect to the Cartesian basis~(\ref{eq19}). The relationships of
amplitudes $\xi (\vec{k})$ with quantum-mechanical
bosonic amplitudes $b (\vec{k})$ of probability
distribution according to the eigenvalues of the stationary
diagonal complete set of operators of quantum-mechanical bosonic
$s=(1,0)$-multiplet are presented by
\begin{equation}
\label{eq21}\xi^{1}=b^{1},\quad \xi^{2}=-\frac{1}{\sqrt{2}}\left(b^{3}+b^{4}\right),\quad \xi^{3}=-\ri b^{2},\quad\xi^{4}=\frac{1}{\sqrt{2}}\left(b^{3}-b^{4}\right);
\quad b^{1,2,3,4}(\vec{k})\equiv
b^{+,-,0,\underline{0}}(\vec{k}),
\end{equation}
where the 4 amplitudes $b^{1,2,3,4}(\vec{k})$
are the quantum-mechanical momentum-spin amplitudes with the eigenvalues $(+1,-1,0,\underline{0})$ of the projection $\breve{s}^{3}$ on the axe 3 of the quantum mechanical spin $s=(1,0)$ multiplet operator $\vec{\breve{s}}$, respectively (the last
eigenvalue \underline{0} is related to the proper zero spin).
Thus, if $\phi_{(1,0)}(x)\in\mathrm{S}^{3,4}$, then the bosonic
amplitudes $\xi (\vec{k})$ belong to the Schwartz
complex-valued test function space too.

Moreover, the set $\{\phi_{(1,0)}(x)\}$ of solutions~(\ref{eq20}) is
invariant, in particular, with respect to the unitary bosonic
representation of the group $\mathcal{P}$, which is determined by
the generators~(\ref{eq16}) and Casimir operators~(\ref{eq17}). Therefore, the
Bargman-Wigner analysis of the Poincar\'{e}
symmetry of the set $\{\phi_{(1,0)}(x)\}$ of solutions~(\ref{eq20})
manifestly demonstrates that this is the set of Bose-states
$\phi_{(1,0)}$ of the field $\phi$, i.e., the $s=(1,0)$-multiplet
states. Hence, the existence of bosonic solutions of the FW
equation is proved.

In terms of quantum-mechanical momentum-spin amplitudes
$b^{\alpha}(\vec{k})$ from~(\ref{eq21}), the bosonic spin (1,0)-multiplet
solution $\psi=V^{+}\phi$ of the Dirac equation~(\ref{eq8}) is presented by
\begin{eqnarray}
\label{eq22}\psi_{(1,0)}(x)&=&\frac{1}{(2\pi)^{{3}/{2}}}\int
\rd^{3}k\Bigg\{\re^{-\ri kx}\left[b^{1}v_{1}^{-}(\vec{k})-\frac{1}{\sqrt{2}}\left(b^{3}+b^{4}\right)
v_{2}^{-}(\vec{k})\right]\nonumber\\
&&+\re^{\ri kx}\left[ib^{*2}v_{1}^{+}(\vec{k})+\frac{1}{\sqrt{2}}\left(b^{*3}-b^{*4}\right)
v_{2}^{+}(\vec{k})\right]\Bigg\} ,
\end{eqnarray}
where the 4-component spinors are the same as in the
Dirac theory of the fermionic doublet
\begin{equation}
\label{eq23}v_{\mathrm{r}}^{-}(\vec{k})=N\left|
{{\begin{array}{*{20}c}
 (\widehat{\omega}+m)\mathrm{d}_{\mathrm{r}}\\
 (\vec{\sigma}\cdot \vec{k})\mathrm{d}_{\mathrm{r}}\\
\end{array} }} \right|,\quad
v_{\mathrm{r}}^{+}(\vec{k})=N\left|
{{\begin{array}{*{20}c}
 (\vec{\sigma}\cdot \vec{k})\mathrm{d}_{\mathrm{r}}\\
 (\widehat{\omega}+m)\mathrm{d}_{\mathrm{r}}\\
\end{array} }} \right|,
\quad N\equiv\frac{1}{\sqrt{2\widehat{\omega}(\widehat{\omega}+m)}}\,,
\quad
\mathrm{d}_{1}=\left| {{\begin{array}{*{20}c}
 1\\
 0\\
\end{array} }} \right|,
\quad
 \mathrm{d}_{2}=\left|
{{\begin{array}{*{20}c}
 0\\
 1\\
\end{array} }} \right|.
\end{equation}

The well known (i.e., standard) Fermi solution of the Dirac equation
for the spin $s=1/2$ doublet has the following form
\begin{equation}
\label{eq24} \psi(x)=\frac{1}{(2\pi)^{{3}/{2}}}\int \rd^{3}k
\left[\re^{-\ri kx}a^{-}_{\mathrm{r}}(\vec{k})v_{\mathrm{r}}^{-}(\vec{k})
+\re^{\ri kx}a^{+}_{\mathrm{r}}(\vec{k})v_{\mathrm{r}}^{+}(\vec{k})\right]
,
\end{equation}
where the physical sense of the amplitudes
$a^{-}_{\mathrm{r}}(\vec{k}),\,a^{+}_{\mathrm{r}}(\vec{k})$
is explained in~\cite{47}.

All the above assertions concerning the FB duality of the spinor
field are valid both in FW and PD representation, i.e., for both
FW~(\ref{eq7}) and Dirac~(\ref{eq8}) equations. The transition between FW and PD
representations is fulfilled by the FW transformation~(\ref{eq6}).

\section{The Fermi-Bose conservation laws for the spinor field}

Let us  briefly note the FB conservation laws (CL) for the spinor
field. It is preferable to calculate them in the FW (nonlocal
PD) representation too. In FW representation, the Fermi spin
$\vec{s}=(s^{23}, s^{31},s^{12})$ from~(\ref{eq11}) (together with the ``boost spin'') is the
independent symmetry operator for the FW equation. The orbital
angular momentum and pure Lorentz angular momentum (the carriers
of external statistical degrees of freedom) are independent symmetry operators in this
representation too (one can also
find the corresponding independent spin and angular momentum
symmetries in the PD representation for the Dirac equation,
but the corresponding operators are essentially nonlocal). Hence,
one obtains 10 Poincar\'{e} and 12 additional (3
spin, 3 pure Lorentz spin, 3 angular momentum, 3 pure angular
momentum)~CL.

Therefore, in the FW representation, one can easily find 22
fermionic and 22 bosonic CL. The division into bosonic and
fermionic set is caused by the existence of FB symmetries and
solutions. Indeed, if substitution of bosonic $\mathcal{P}$
generators $q$~(\ref{eq16}) and bosonic solutions~(\ref{eq20}) into the Noether
formula $Q=\int \rd^{3}x\phi^{\dag}(x) q\phi(x)$ is made, then
automatically the bosonic CL for $s=(1,0)$-multiplet are obtained.
The standard substitution of the corresponding well known fermionic
generators and solutions presents fermionic~CL.

We briefly illustrate the difference in fermionic and bosonic CL based
on the example of the corresponding spin conservation. For a
fermionic spin
\begin{equation}
\label{eq25} \vec{s}=
(s_{23},s_{31},s_{12})\equiv(s^{\ell})= \frac{1}{2}\left|
{{\begin{array}{*{20}c}
 \vec{\sigma} \hfill & 0 \hfill \\
 0 \hfill &  \vec{\sigma} \hfill \\
\end{array} }} \right|\rightarrow s_{z}\equiv s^{3}=\frac{1}{2}\left|
\begin{array}{cccc}
 1 & 0 & 0 & 0\\
 0 & -1 & 0 & 0\\
0 & 0 & 1 & 0\\
0 & 0 & 0 & -1\\
\end{array} \right|
\end{equation}
and for a bosonic spin~(\ref{eq12}), (\ref{eq14}), the CL are given by
\begin{equation}
\label{eq26} S_{mn}^{\mathrm{F}}=\int
\rd^{3}x\phi^{\dag}(x) s_{mn}\phi(x)=\int
\rd^{3}kA^{\dag}(\vec{k}) s_{mn}A(\vec{k}),
\end{equation}
\begin{equation}
\label{eq27} S_{mn}^{\mathrm{B}}=\int
\rd^{3}x\phi^{\dag}(x)\breve{s}_{mn}\phi(x)=\int
\rd^{3}kB^{\dag}(\vec{k})\breve{s}_{mn}B(\vec{k}),
\end{equation}
where
\begin{equation}
\label{eq28}
A(\vec{k})=\mathrm{column}\left(a^{-}_{+},\,a^{-}_{-},\,a^{*+}_{-},\,a^{*+}_{+}\right),
\qquad
B(\vec{k})=\mathrm{column}\left(b^{1},\,b^{2},\,b^{*3},\,b^{*4}\right).
\end{equation}

We present these CL in terms of quantum-mechanical Fermi and Bose
amplitudes. All integral conserved quantities have an explicit quantum-statistical form.

\section{The Fermi-Bose duality and the Fermi-Bose statistics}

An adequate statistical quantum-mechanical sense of the coefficients
$a^{-}_{\mathrm{r}}(\vec{k})$, $a^{+}_{\mathrm{r}}(\vec{k})$
in the expansion~(\ref{eq24}) over the basis solutions~(\ref{eq23}) of the Dirac
equation is found in the same way but with the help of transition
$\phi(x)=V^{-}\psi(x)$~(\ref{eq9}), (\ref{eq6})  to the FW representation~\cite{43}.
Indeed, the statistical sense of the FW field $\phi(x)$ is
evidently related to the statistical sense of the
particle-antiparticle doublet in relativistic canonical quantum
mechanics~\cite{43,48} of this doublet. It is shown in~\cite{43} that
\begin{equation}
\label{eq29} \phi=\left| {{\begin{array}{*{20}c}
 \phi^{-}\\
 0\\
\end{array} }} \right|+\left|
{{\begin{array}{*{20}c}
 0\\
 \phi^{*+}\\
\end{array} }} \right|,
\end{equation}
where $\phi^{\mp}(x)$ are the relativistic
quantum-mechanical wave functions of the particle-antiparticle
doublet.

The solution of the FW equation~(\ref{eq7}) expanded over the eigenvectors
of quantum-mechanical fermi\-onic stationary diagonal complete set
of operators  (momentum $\vec{p}$, projection $s^{3}$
of the spin $\vec{s}^{\mathrm{quant.-mech.}}$ and sign
of the charge $g=-\gamma^{0}$) has the following form
\begin{equation}
\label{eq30} \phi(x)=\frac{1}{(2\pi)^{{3}/{2}}}\int
\rd^{3}k\left\{\re^{-\ri kx}\left[a^{-}_{+}(\vec{k})\mathrm{d}_{1}+a^{-}_{-}(\vec{k})\mathrm{d}_{2}\right]
+\re^{\ri kx}\left[a^{*+}_{-}(\vec{k})\mathrm{d}_{3}+a^{*+}_{+}(\vec{k})\mathrm{d}_{4}\right]\right\}
,
\end{equation}
where the coefficients of expansion
$a^{-}_{+}(\vec{k})$, $a^{-}_{-}(\vec{k})$,
$a^{+}_{-}(\vec{k})$, $a^{+}_{+}(\vec{k})$
denote the statistical quantum-mechanical amplitudes of
probability distribution over the eigenvalues of the above
mentioned fermionic stationary complete set of operators. The
4-columns $\mathrm{d}_{\alpha}=(\delta_{\alpha}^{\beta})$ are the
Cartesian orts in the space
$\mathrm{C}^{\otimes4}\subset\mathrm{H}^{3,4}$. In order to obtain
the most adequate and obvious statistical quantum-mechanical
interpretation of the amplitudes and solutions, the spin
projection operator in a complete set (momentum
$\vec{p}$, projection $s^{3}$ of the spin
$\vec{s}^{\mathrm{quant.-mech.}}$ and sign of the
charge $g=-\gamma^{0}$) is taken in the quantum-mechanical form~\cite{48}
\begin{equation}
\label{eq31} \vec{s}^{\mathrm{quant.-mech.}}=
\frac{1}{2}\left| {{\begin{array}{*{20}c}
 \vec{\sigma} \hfill  0 \hfill \\
 0 -C\hfill\vec{\sigma}\hfill C \\
\end{array} }} \right|\rightarrow s_{z}^{\mathrm{quant.-mech.}}\equiv s^{3}=\frac{1}{2}\left|
\begin{array}{cccc}
 1 & 0 & 0 & 0\\
 0 & -1 & 0 & 0\\
0 & 0 & -1 & 0\\
0 & 0 & 0 & 1\\
\end{array} \right|
\end{equation}
rather than in the canonical field theory form~(\ref{eq25}). The
statistical sense of the amplitudes is conserved in the solution
[$\psi(x)=V^{+}\phi(x)$~(\ref{eq9}), (\ref{eq6})]
\begin{eqnarray}
\label{eq32} \psi(x)&=&\frac{1}{(2\pi)^{{3}/{2}}}\int
\rd^{3}k\Big\{\re^{-\ri kx}\left[a^{-}_{+}(\vec{k})v_{1}^{-}(\vec{k})+a^{-}_{-}(\vec{k})
v_{2}^{-}(\vec{k})\right]\nonumber\\
&&+\re^{\ri kx}\left[a^{*+}_{-}(\vec{k})v_{1}^{+}(\vec{k})+a^{*+}_{+}(\vec{k})
v_{2}^{+}(\vec{k})\right]\Big\}
\end{eqnarray}
of the Dirac equation~(\ref{eq8}) in its standard local
representation. The amplitudes
$a^{-}_{+}(\vec{k})$, $a^{-}_{-}(\vec{k})$, $a^{+}_{-}(\vec{k})$, $a^{+}_{+}(\vec{k})$
in the fermionic solutions~(\ref{eq30}) and~(\ref{eq32}) of the FW and Dirac
equations are one and the same. Thus,
$a^{-}_{+}(\vec{k})$, $a^{-}_{-}(\vec{k})$
are the quantum-mechanical momentum-spin amplitudes of the
particle with charge $-e$ and eigenvalues of spin projection $+1/2$
and $-1/2$;
$a^{+}_{-}(\vec{k})$, $a^{+}_{+}(\vec{k})$
are the quantum-mechanical momentum-spin amplitudes of the
antiparticle with charge $+e$ and eigenvalues of spin projection
$-1/2$ and $+1/2$, respectively.

Statistical quantum mechanical sense of the bosonic amplitudes
$b^{\alpha}(\vec{k})$ of bosonic solution~(\ref{eq22}) of the
Dirac equation~(\ref{eq8}) is found in the same way and is explained in
section~5 in the process of constructing this solution.

The relationship between the fermionic
$a^{-}_{+}\!(\vec{k})$, $a^{-}_{-}\!(\vec{k})$, $a^{+}_{-}\!(\vec{k})$, $a^{+}_{+}\!(\vec{k})$
and bosonic $b^{1,2,3,4}(\vec{k})\equiv
b^{+,-,0,\underline{0}}(\vec{k})$ amplitudes in one and
the same (arbitrarily fixed) physical state of FB dual field
$\psi$ is presented by the unitary operator $U$ in the following form:
\begin{equation}
\label{eq33} \left|
\begin{array}{cccc}
 a^{-}_{+} \\
 a^{-}_{-} \\
 a^{+}_{-} \\
 a^{+}_{+} \\
\end{array} \right|=\frac{1}{\sqrt{2}}\left|
\begin{array}{cccc}
 \sqrt{2} & 0 & 0 & 0\\
 0 & 0 & -1 & -1\\
0 & -\ri\sqrt{2} & 0 & 0\\
0 & 0 & 1 & -1\\
\end{array} \right|\left|
\begin{array}{cccc}
 b^{+} \\
 b^{-} \\
 b^{0} \\
 b^{\underline{0}} \\
\end{array} \right|,\qquad \left|
\begin{array}{cccc}
 b^{+} \\
 b^{-} \\
 b^{0} \\
 b^{\underline{0}} \\
\end{array} \right|=\frac{1}{\sqrt{2}}\left|
\begin{array}{cccc}
 \sqrt{2} & 0 & 0 & 0\\
 0 & 0 & \ri\sqrt{2} & 0\\
0 & -1 & 0 & 1\\
0 & -1 & 0 & -1\\
\end{array} \right|\left|
\begin{array}{cccc}
 a^{-}_{+} \\
 a^{-}_{-} \\
 a^{+}_{-} \\
 a^{+}_{+} \\
\end{array} \right|.
\end{equation}
Relationships~(\ref{eq33}) follow directly from the comparison
of the solutions~(\ref{eq22}) and~(\ref{eq32}).

Note that the set of fermionic solutions $\{\psi^{\mathrm{F}}\}$~(\ref{eq32})
of the Dirac equation is invariant with respect to the well
known induced fermionic $\mathcal{P}^{\mathrm{F}}$ representation
of the Poincar\'{e} group $\mathcal{P}$~\cite{49}, see
also formula (19) in the paper~\cite{41}). The set of bosonic solutions
$\{\psi^{\mathrm{B}}\}$~(\ref{eq22}) of the Dirac equation is invariant
with respect to the induced bosonic $\mathcal{P}^{\mathrm{B}}$
representation of the Poincar\'{e} group
$\mathcal{P}$ (formula (21) in the paper~\cite{42}). However, the
relationships~(\ref{eq33}) between the fermionic
$a^{-}_{+}(\vec{k})$, $a^{-}_{-}(\vec{k})$,
$a^{+}_{-}(\vec{k})$, $a^{+}_{+}(\vec{k})$
and bosonic $b^{1,2,3,4}(\vec{k})\equiv
b^{+,-,0,\underline{0}}(\vec{k})$ amplitudes do not
change in any inertial frame of references.

\section{Conclusions}

The 64 dimensional ERCD and 29 dimensional proper ERCD algebras,
which have been put into consideration in~\cite{40,41,42}, are  useful
generalizations of the standard 16 dimensional CD algebra. Application thereof enabled us to prove the existence of additional
bosonic symmetries, solutions and conservation laws for the
spinor field, the Foldy-Wouthuysen and the Dirac equations. The
investigation of the spinor field in the Foldy-Wouthuysen
representation has the sense and purpose of its own. This
representation is of great interest in itself, especially in connection with the recent
result~\cite{50} by V.~Neznamov, who developed the formalism of quantum
electrodynamics in the Foldy-Wouthuysen representation, see also
the results in~\cite{51}. The property of the Fermi-Bose duality of
the Dirac equation (both in the Foldy-Wouthuysen and the
Pauli-Dirac representations), which proof was started in~\cite{39,40,41,42},
where the bosonic symmetries of this equation were found, is
demonstrated herein on the next level, i.e., on the level of the existence
of the spin (1,0) bosonic solutions of the equation under
consideration and corresponding bosonic conservation laws.
Similarly, the fermionic spin $s=1/2$ properties for the Maxwell
equations both with nonzero and zero mass can be proved (see e.g., the procedure given in~\cite{6}).

In any case, we do not change the main well known postulates and
theory of the Fermi-Bose statistics. Our results have another,
principally new  meaning. In our approach, the Fermi-Bose duality
of the spinor field found in~\cite{2} is proved based on the examples of
the existence of bosonic symmetries (section~4) and solutions
(section~5) of the Dirac equation with nonzero mass together with
the confirmation of the bosonic conservation laws (section~6) for the spinor
field. This opens up new possibilities for applying the Dirac equation
to the description of bosonic states. Thus, the
property of the Fermi-Bose duality of the Dirac equation that was proved
in our publications~\cite{39,40,41,42} and in the present paper does not break the Fermi
statistics for fermions (with the Pauli principle) and Bose
statistics for bosons (with Bose condensation). We have never
mixed up the Fermi and Bose statistics as well. Our
assertion is as follows. One can apply with equal success both
Fermi and Bose statistics for one and the same Dirac equation and
for one and the same spinor field, i.e., the Dirac equation can
describe both fermionic and bosonic states.

\ukrainianpart

\title{Деякі статистичні аспекти Фермі-Бозе дуалізму \\ спінорного поля}
\author{В.М. Симулик, І.Ю. Кривський, І.Л. Ламер}
\address{
Інститут електронної фізики, Національна академія наук України, \\ вул. Університетська, 21, 88000 Ужгород,
Україна}

\makeukrtitle

\begin{abstract}
\tolerance=3000%
Коротко розглядається структура 29-вимірної розширеної дійсної
алгебри Кліффорда-Дірака, яка була введена в розгляд у нашій
публікації Phys. Lett. A, 2011,
\textbf{375}, 2479. На основі цієї алгебри доводиться
властивість Фермі-Бозе дуалізму рівняння Дірака з ненульовою
масою. Це означає, що рівняння Дірака може описувати не лише
ферміонні, але й бозонні стани. Доведення дається на прикладах
наявності бозонних симетрій, розв'язків та законів збереження.
Розглянуто деякі статистичні аспекти Фермі-Бозе дуалізму
спінорного поля

\keywords спінорне поле, симетрія, теоретико-груповий аналіз,
суперсиметрія, представлення Фолді-Ваутхайзена, алгебра Кліффорда-Дірака, Фермі-Бозе дуалізм

\end{abstract}

\end{document}